# Title: Antiferromagnet-based spintronic functionality by controlling isospin domains in a layered perovskite iridate

*Nara Lee, Eunjung Ko, Hwan Young Choi, Yun Jeong Hong, Muhammad Nauman, Woun Kang, Hyoung Joon Choi\*, Young Jai Choi\*, and Younjung Jo\**

Dr. N. Lee, Dr. E. Ko, H. Y. Choi, Prof. H. J. Choi, Prof. Y. J. Choi
Department of Physics,
Yonsei University,
Seoul 03722, Korea
E-mail: h.j.choi@yonsei.ac.kr and phylove@yonsei.ac.kr
Y. J. Hong, M. Nauman, Prof. Y. Jo
Department of Physics,
Kyungpook National University,
Daegu 41566, Korea
E-mail: jophy@knu.ac.kr
Prof. W. Kang
Department of Physics,
Ewha Womans University,
Seoul 03760, Korea

Abstract: The novel electronic state of the canted antiferromagnetic (AFM) insulator, strontium iridate ($Sr_2IrO_4$) has been well described by the spin-orbit-entangled isospin $J_{eff} = 1/2$, but the role of isospin in transport phenomena remains poorly understood. In this study, antiferromagnet-based spintronic functionality is demonstrated by combining unique characteristics of the isospin state in $Sr_2IrO_4$. Based on magnetic and transport measurements, large and highly anisotropic magnetoresistance (AMR) is obtained by manipulating the antiferromagnetic isospin domains. First-principles calculations suggest that electrons whose isospin directions are strongly coupled to in-plane net magnetic moment encounter the isospin mismatch when moving across antiferromagnetic domain boundaries, which generates a high resistance state. By rotating a magnetic field that aligns in-plane net moments and removes domain boundaries, the macroscopically-ordered isospins govern dynamic transport through the system, which leads to the extremely angle-sensitive AMR. As with this work that establishes a link between isospins and magnetotransport in strongly spin-orbit-coupled AFM

Sr$_2$IrO$_4$, the peculiar AMR effect provides a beneficial foundation for fundamental and applied research on AFM spintronics.

Antiferromagnetic (AFM) spintronics is an emerging field that aims to achieve spintronic functionality by taking advantages of AFM materials such as no stray fields, insensitivity to disturbing magnetic fields, and intrinsically-fast AFM dynamics[1]. The main issue to generate functional spin transport based on antiferromagnets is to manipulate and detect an AFM state. The AFM memory state has been examined in several studies using anisotropic magnetoresistance (AMR)[2-4]; however, it mostly requires complicated stacking geometry for unified spintronic functionality. Since the AFM metallic alloys were mostly adopted for the desired functionality[5-10], AMR effects in ohmic magnetic devices are ordinarily restricted to variation of a few percent[11], which motivated recent studies focusing on tunnelling AMR[5, 12-14]. Due to the transport of an AFM tunnelling junction governed by tunnelling probability through an insulating layer, an enhanced MR effect is attained in favour of high density MRAMs[2, 5].

In the canted AFM Mott insulator Sr$_2$IrO$_4$, the nontrivial quantum state carries a magnetic moment described by the spin-orbit-entangled isospin $J_{eff} = 1/2$, which fully expresses the relativistic spin-orbit coupling under a large crystal field. The combined action of Coulomb interactions and AFM order results in the further splitting of the $J_{eff} = 1/2$ band, thus opening a gap[15-17]. Despite the identical formalism with a conventional $J = 1/2$ quantum state, the $J_{eff} = 1/2$ isospin state is distinct because it is formed by the combination of spatially anisotropic orbitals and pure spins. While the unconventional insulating state in Sr$_2$IrO$_4$ is well described by the $J_{eff} = 1/2$ state, the importance of correlation between the isospin state and electronic transport behavior has scarcely been appreciated. Moreover, recent investigations have suggested that spin-dependent transport in Sr$_2$IrO$_4$ is fundamental for new antiferromagnet-based spintronics[18], and have stressed the need to understand the impact of the isospin state on magnetotransport[2, 10, 19]. Here we demonstrate a large and highly anisotropic negative magnetoresistance that produces negligible stray fields in single crystalline Sr$_2$IrO$_4$. Due to the unique aspect of the canted antiferromagnetism in Sr$_2$IrO$_4$, the macroscopic AFM order of isospins is controlled by the intrinsic weak-ferromagnetic moment, which governs dynamic transport and leads to the extremely angle-sensitive large AMR. The AFM spintronic functionality by integrating intrinsic bulk properties of a single-phase material simplifies the stacking geometry and thus offers a more expandable platform to construct sophisticated heterostructures.

Sr$_2$IrO$_4$ crystallizes in a layered perovskite structure (tetragonal, $I4_1/acd$) with four-fold rotational symmetry ($C_4$) about the [001] axis[20]. **Figure 1**a displays the temperature dependence of resistivity measured in a zero magnetic field with current along the [$\bar{1}$10] direction. The electronic transport is promoted by bulk electron doping resulting from proper oxygen deficiency using the benefit of efficiently controlled oxidation in an oxide. The temperature-dependent magnetic susceptibility was measured upon warming at 0.2 T after zero-field cooling and upon cooling at 0.2 T for $H_{90}$ and $H_0$ (Figure 1b), where $H_\theta$ is the magnetic field deviating from the [001] axis by the angle $\theta$ ($\theta = 0°$ for the [001] direction and $\theta = 90°$ for the [110] direction) (inset of Figure 1a). As the temperature decreases, the susceptibility for $H_{90}$ increases abruptly at $T_N \approx 220$ K signifying the emergence of canted AFM order, while the resistivity does not present any anomaly at $T_N$, consistent with the previous results[21, 22]. A canted antiferromagnet, known alternatively as a weak ferromagnet, contains both antiferro- and ferro-magnetic characteristics. Though it exhibits antiparallel alignment of microscopic magnetic moments as in antiferromagnetism, it generates a small net magnetization and follows domain processes as in ferromagnetism due to slight canting of magnetic moments. The susceptibilities for two different orientations indicate considerable magnetic anisotropy associated with the isospins mainly aligned in the *ab* plane. Accompanied by the oxygen deficiency in crystals, $T_N$ is lowered, but magnetic anisotropy remains intact.

To investigate the effect of the macroscopic ordering of isospins on transport, we measured magnetoresistance (MR) in $H_\theta$, i.e., MR$_\theta$ (%) = $\frac{R(H_\theta) - R(0\ \text{T})}{R(0\ \text{T})} \times 100$ (1), where $R$ denotes the resistance and $H_\theta$ was applied perpendicularly to the current (inset of Figure 1a), precluding the ordinary Lorentzian MR effect, and also measured the isothermal magnetization in $H_{90}$ ($M_{90}$) on a single Sr$_2$IrO$_4$ crystal. The MR$_{90}$ at 30 K decreases substantially with increasing $H_{90}$ and almost saturates above $H_{90} \approx 0.3$ T with a large negative value of ~33%, as shown in Figure 1c. The consecutive sweeping of $H_{90}$ between 0.8 and –0.8 T gives rise to slight hysteresis. Consistently, the $M_{90}$ at 30 K shows tiny magnetic hysteresis and saturation above $H_{90} \approx 0.3$ T with ~0.04 $\mu_B$/f.u. (Figure 1d). This strong relationship between the MR$_{90}$ and $M_{90}$ data indicates that isospin ordering controls magnetotransport in the system.

Below $T_N$, the isospins exhibit a planar canted AFM order owing to the staggered rotation of the IrO$_6$ octahedra[23]. Concurrently, the $C_4$ symmetry is lowered to $C_2$ (two-fold rotational symmetry), producing an orthorhombic magnetic structure[23-25]. Therefore, the four types of magnetic domain are populated equally in zero magnetic field (see Section S2, Supporting Information). In a magnetic unit cell, the net magnetic moment ($\mu_{net}$) in each IrO$_2$ layer of the unit cell is stacked along the [001] axis in a way that $\mu_{net}$'s of consecutive IrO$_2$ layers are cancelled out. When a magnetic field above the critical strength ($H_{90} \approx 0.3$ T) is applied in the *ab* plane, a single weak ferromagnetic state is formed because of the alignment of $\mu_{net}$'s (see Section S2, Supporting Information). In Sr$_2$IrO$_4$, the susceptible response of $\mu_{net}$'s to the external magnetic field leads to a narrow magnetic hysteresis loop with negligible amount of residual net magnetization (Figure 1d), similar to the behavior of a soft ferromagnet. In such a process, the high resistance state with four types of magnetic domains evolves to the low resistance state with a single weak-ferromagnetic domain by applying a magnetic field in the *ab* plane. The almost reversible magnetization process recovers the high resistance state once the magnetic field is removed.

In contrast to MR$_{90}$, MR$_0$ decreases gradually with increasing $H_0$ and reveals a small variation of ~3% between 0 and 4 T (Figure 1e), which reflects the strong magnetic anisotropy. A slight deviation of $\theta$ by 1° reduces the resistance noticeably, inducing a change of ~20% in the range 0–4 T. With further rotation of $H_\theta$ beyond $\theta = 5°$, MR$_\theta$ drops precipitously and reaches the maximum variation within 2 T. Figure 1f presents the MR$_\theta$ data of Figure 1e plotted with respect to the effective magnetic field ($H_{\text{eff}}$), defined as the projection of the $H_\theta$ onto the [110] axis, i.e., $H_{\text{eff}} = H_\theta |\sin\theta|$. All the MR$_\theta$ curves overlap with MR$_{90}$, suggesting that the magnetic transport is dominated solely by the in-plane component of the applied magnetic field.

The MR$_{90}$ behavior that is strongly relevant to the saturation of $M_{90}$ by aligning the $\mu_{\text{net}}$'s implies that the magnetotransport would be caused by magnetic-field-dependent isospin structure rather than by electronic band dispersion. To validate the assertion, we have calculated the electronic band structures (see Section S2, Supporting Information) corresponding to the high- and low-resistance states. It appears that the band structure representing the single weakly ferromagnetic state shows that the band gap is 0.0074 eV smaller than that in one of the four types of magnetic domain. Therefore, the result justifies that the MR$_{90}$ behavior would not be explained by the conventional electronic band model. The intimate correlation between MR$_{90}$ and $M_{90}$ was examined by performing first-principles calculations of the isospin structure for magnetic domains of different directions of $\mu_{\text{net}}$ in Sr$_2$IrO$_4$ (see Section S3 for detail, Supporting Information). As the electronic-band structure of Sr$_2$IrO$_4$ (**Figure 2**a) is close to a four-fold rotational symmetry with respect to the [001] axis, the in-plane rotation of $\mu_{net}$ by 90° or 180° does not change the band dispersion significantly (see Section S2, Supporting Information). However, we find that the isospin directions near points X-S, which connect the conduction band minimum (CBM) at X with the valence band maximum (VBM) at S (Figure 2a), are strongly coupled to the direction of $\mu_{net}$ (see Section S3 for detail, Supporting Information). The isospin vectors of the conduction band (CB) states at a crystal-momentum grid before (red arrows) and after rotation of the $\mu_{net}$ (blue arrows) by 90° and 180° are shown in Figure 2b and 2c, respectively. The vectors can be determined by calculating the expectation value of the $J_{eff}$ operator with respect to the CB states. By comparison of the CB states before and after a 90°-rotation of the $\mu_{net}$ near points X-S, the squared inner products of the quantum-mechanical wavefunctions including spin states are estimated as 0.5 (Figure 2d), which signifies the 50% transmission of the current. In the case of a 180° rotation of $\mu_{net}$, the CB states (Figure 2e) before and after the rotation are orthogonal to each other, that is, their squared inner products are zero. Similar results are obtained for the valence-band (VB) states (see Section S3 for detail, Supporting Information). We note that the squared inner products are sensitive to the rotation of $\mu_{net}$ near the X-S line because the isospin vector is an effective quantum number for CB states near the X-S line.

In Figure 2f and 2g, the isospins of VB states, which have directions opposite to those of CB states (see Section S3 for detail, Supporting Information), are illustrated as arrows onto the $Ir^{4+}$ ions for two possible types of domain boundaries with 90° (f) and 180° (g) difference in the direction of $\mu_{net}$, respectively. The translucent red arrows represent partial (f) and complete (g) reflection of the electrical current at domain boundaries, illustrating the result of squared inner products. The estimated strong dependence of CB and VB states on the direction of $\mu_{net}$ provides a microscopic picture of the observed $MR_{90}$. Without an applied magnetic field, a $Sr_2IrO_4$ sample consists of four different types of magnetic domains, having zero total magnetization of the sample. In this case, CBM and VBM electrons moving across the domain boundaries experience isospin mismatch, which generates scattering or reflection that increases the electrical resistance. When an applied magnetic field aligns $\mu_{net}$'s and removes the domain boundaries, the scattering at the domain boundaries disappears, creating a negative $MR_{90}$ that saturates when the $M_{90}$ is saturated.

The strongly anisotropic isospin-dependent magnetotransport results in the exceedingly sensitive angular dependence of AMR, obviously revealed in the crystalline component that is relevant entirely to the crystal symmetry (**Figure 3**a). The AMR, defined as AMR (%) = $\frac{R(\theta)-R(0°)}{R(0°)} \times 100$  (2), was measured by rotating $H_\theta$ = 0.5 T in the plane perpendicular to the current (inset of Figure 1a) at different temperatures. A small AMR effect, which is comparable to the typical AMR described by a $\sin^2\theta$ dependence[26], emerges at 210 K. Upon cooling further, the negative AMR effect increases with decreasing temperature. A slight deviation of $\theta$ from the [001] axis quickly reduces the AMR and further rotation leads to little variation with the maximum value achieved at $\theta$ = 90°. Note that the strong temperature dependence of the AMR was examined by devising a microscopic theory for the electron transport (see Section S4 for detail, Supporting Information). The feature of the unconventional angular dependence of AMR is plotted evidently in polar angular dependence measured at 30 K (Figure 3b) with the maximum magnitude of AMR as ~33%, which is differentiated from the conventional sinusoidal variation of the AMR obtained by rotating a magnetic field in the *ab* plane in the previous works (see Section S5 for detail, Supporting Information) [19, 27, 28]. Figure 3c shows the complete AMR contour map measured at 30 K for $H_\theta$ between −1 T and 4 T and $\theta$ between −90° and 90°. With increasing $H_\theta$, the susceptible angular dependence near $\theta$ = 0° becomes extremely sharpened owing to the incorporation of the in-plane magnetic field component within a smaller angle variation.

Magnetic anisotropy is ubiquitous in single-crystalline magnetic materials. However, not many of them reveal the MR or crystalline AMR effect. In $Sr_2IrO_4$, the magnetically anisotropic $J_{eff}$ = 1/2 state plays a central role in the large and angle-sensitive AMR effect. As the direction of the $\mu_{net}$ determines the arrangement of isospins, it has a strong influence on the character of the wave function. In detail, when the $\mu_{net}$ is rotated by 180°, all the isospins within the $IrO_2$ plane are also rotated by 180°. This collective 180°-rotation of the isospins corresponds to performing time-reversal-symmetry transformation of the wave function in the conduction bands. Since a wave function after the time-reversal symmetry transformation is completely orthogonal to that before the transformation, the 180°-rotation of the $\mu_{net}$ changes the wave function to an orthogonal one. As a result, the squared inner products of wave functions before and after the rotation leads to the complete reflection of the electrical current at the domain boundary. Similarly, the 90°-rotation of $\mu_{net}$ results in the 50 % transmission of the current through the boundary. The four types of magnetic domain are maintained when $H$ = 0.5 T is applied to the [001] axis because of the strong magnetic anisotropy and the isospin mismatch across the domain boundaries increases the electrical resistance. The slight deviation of the magnetic field from the [001] axis including the in-plane magnetic component above the critical strength induces the single weakly ferromagnetic state by removing the domain boundaries and the angle-sensitive large negative AMR effect of ~33%.

The repeatability of the modulation of the AMR value was demonstrated by varying $\theta$ linearly between −7.5° and 7.5° at $H_\theta$ = 1 T and 30 K, as shown in Figure 3d. Starting at $\theta$ = 0°, a slight positive rotation of $H_\theta$ rapidly reduces the resistance, and the total variation of AMR up to $\theta$ = 7.5° is approximately −30%. As the angle decreases to $\theta$ = 0°, the initial resistance value is recovered. Subsequent rotation of $H_\theta$ in the opposite direction generates the same effect. The large sequential oscillatory response of AMR to small angle variations is maintained with no significant decay. A similar AMR variation of ~1.5% is observed at 190 K for an angle oscillation between −3.0° and 3.0° (Figure 3e). Such sensitive modification of the AMR in small angular ranges may provide practical benefits for applications in highly sensitive AMR sensor technology.

In our studies, we achieved the antiferromagnet-based spintronic functionality in an insulating oxide material of canted AFM $Sr_2IrO_4$. The unique insulating state in $Sr_2IrO_4$ described by the spin-orbit-entangled $J_{eff}$ = 1/2 state enables us to use a different experimental approach. The insulating gap was modified by the oxygen deficiency inherent from the growth conditions promoting charge transport, which may be essential for incorporating semiconducting and spintronic performance. Different from conventional electronic transport, each charge carrier in $Sr_2IrO_4$ contains the well-defined isospin whose direction is controlled by the net magnetic moment. Consequently, the isospin mismatch across the domain boundaries acts as an essential mechanism for the angle-sensitive large negative AMR effect of ~33%. In consideration of recent advances that investigate novel quantum states and relevant functional properties in strongly spin-orbit-coupled materials, the antiferromagnet-based isospintronics in $Sr_2IrO_4$ is closely relevant to ongoing efforts to achieve improved AFM devices.

**Experimental Section**

*Growth of $Sr_2IrO_4$ single crystals*: We synthesized single crystals of $Sr_2IrO_4$ by utilizing a flux method with $SrCl_2·6H_2O$ flux in air. A mixture of pre-sintered $Sr_2IrO_4$ polycrystalline powder and flux with 1:15 ratio was heated to 1,380 °C in a Pt crucible, dwelled at 1,380 °C for 5 hours, cooled slowly at 8 °C/h to 900 °C, and then cooled in a furnace after the power was turned off.

*Magnetic and transport property measurements*: Temperature and magnetic field dependence of magnetic properties were measured using a SQUID magnetometer (Magnetic Properties

Measurement System (MPMS), Quantum Design Inc. (San Diego, CA, USA)). Electrical transport properties were measured using the conventional four-probe configuration in a Physical Properties Measurement System (PPMS) (Quantum Design Inc.). $MR_\theta$ and AMR at various magnetic fields and different temperatures were measured with a single-axis rotator by polar angle scan of magnetic field.

*Electronic structure calculations*: The electronic structure of $Sr_2IrO_4$ was calculated using noncollinear spin-dependent density functional theory (DFT) with the LDA to the exchange-correlation energy and *ab-initio* norm-conserving pseudopotentials by using SIESTA code[29]. The on-site Coulomb interactions between Ir *d*-orbitals were considered using the rotationally invariant form of the DFT+*U* method, with Coulomb parameter $U$ = 2.54 eV and Hund's parameter $J$ = 0.60 eV which were obtained from the constrained LDA method[30]. The spin-orbit interaction was evaluated using fully relativistic pseudopotentials. The electron density was obtained by integrating the wave functions with a 8 × 8 × 8 *k*-grid in the full Brillouin zone. Real-space grids were generated with a cutoff energy of 1500 Ry to represent the electron density distribution. A unit cell containing 56 ions including eight $Ir^{4+}$ ions was used with experimental lattice constants of *a* = 5.48463Å and *c* = 25.7977Å for the electronic structure calculation in the canted AFM phase.

**Supporting Information**

Supporting Information is available from the Wiley Online Library or from the author.

**Acknowledgements**

This work was supported by the NRF Grant (2011-0018306, NRF-2013R1A1A2063904, 2015-001948, NRF-2016R1C1B2013709, NRF-2016R1A2B4016656, NRF-2017K2A9A2A08000278, 2017R1A5A1014862 (SRC program: vdWMRC center), and NRF-

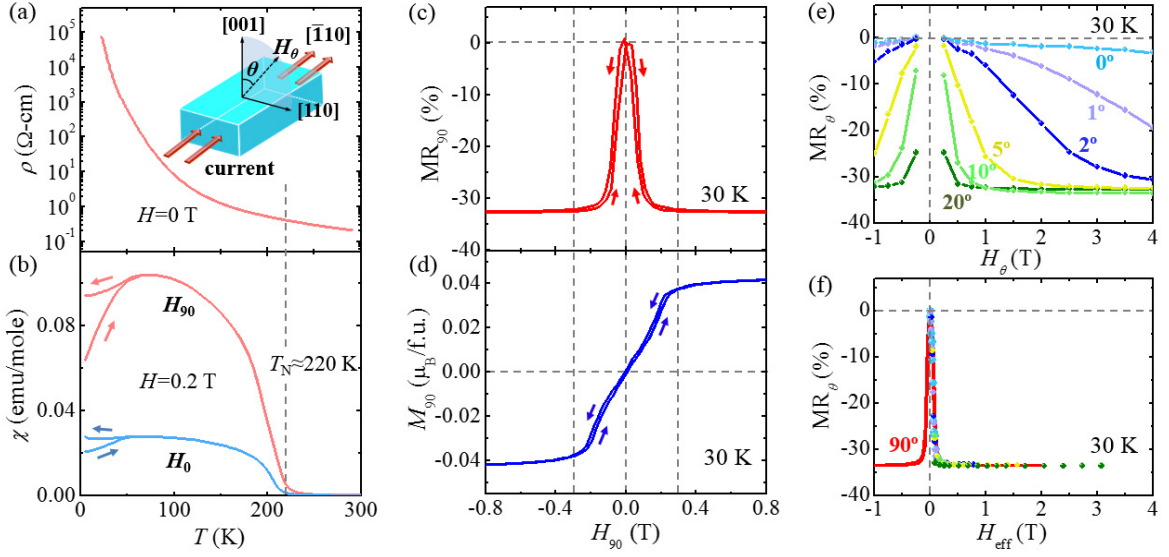

**Figure 1.** (Color online). (a) Temperature dependence of resistivity measured with current along the [$\bar{1}$10] direction at zero magnetic field. The inset shows the geometry of the magnetic field for magnetotransport and magnetization measurements. (b) Temperature dependence of the magnetic susceptibility $\chi=M/H$, measured upon warming at 0.2 T after zero-field cooling and upon cooling at 0.2 T for $H_{90}$ (light red) and $H_0$ (light blue), where $H_\theta$ is the magnetic field deviating from the [001] axis by $\theta$. (c) MR in $H_{90}$ (MR$_{90}$) measured using positive and negative sweeps of the $H_{90}$ up to ±0.8 T at 30 K. (d) Isothermal magnetization along the [110] direction ($M_{90}$), measured using positive and negative sweeps of the magnetic field up to ±0.8 T at 30 K. (e) MR at different angles ($\theta = 0-20°$), MR$_\theta$, plotted in $H_\theta$ between –1 T and 4 T at 30 K. (f) The MR$_\theta$ values of (e) expressed as a function of the effective magnetic field ($H_{\text{eff}}$), i.e., the projection of the applied magnetic field onto the [110] direction, in comparison with MR$_{90}$ in (c).
14

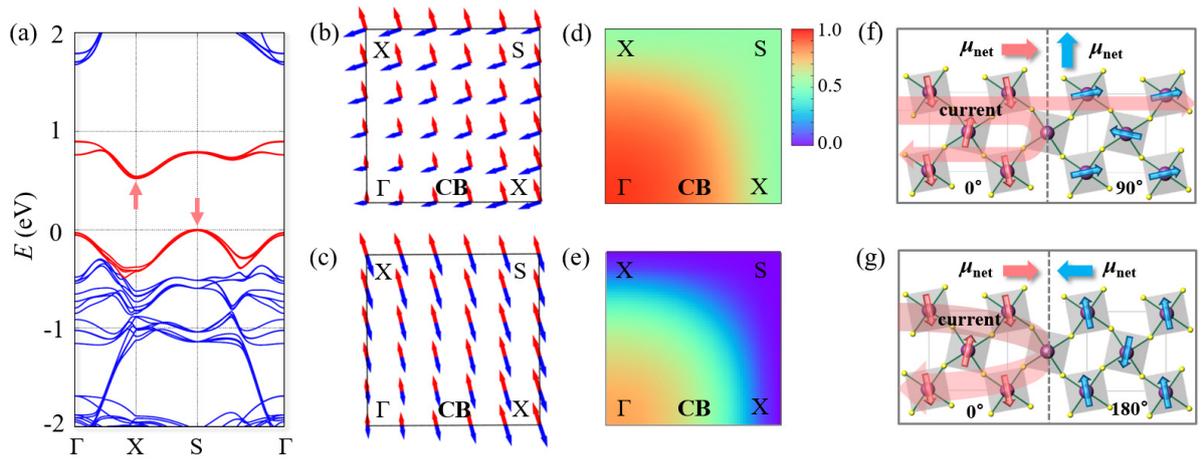

**Figure 2.** (Color online). (a) Band structure of the noncollinear local spin density approximation plus U (LSDA+U) method with SOC. The arrows indicate the CBM (X) and the VBM (S). (b)-(c) Isospin vectors of CB states at a crystal-momentum grid before and after rotation of $\mu_{net}$ by 90° (b) and 180° (c). In (b) and (c), isospin vectors, shown in red and blue arrows before and after the rotation, respectively, rotate by the same angle as $\mu_{net}$. (d)-(e) Squared inner products of CB states at the crystal-momentum grid before and after the rotation of $\mu_{net}$ by 90° (d) and 180° (e). (f)-(g) Schematic representations of partial (f) and complete (g) reflection of the electrical current (translucent red arrows) at domain boundaries with 90° (f) and 180° (g) difference in the direction of $\mu_{net}$.

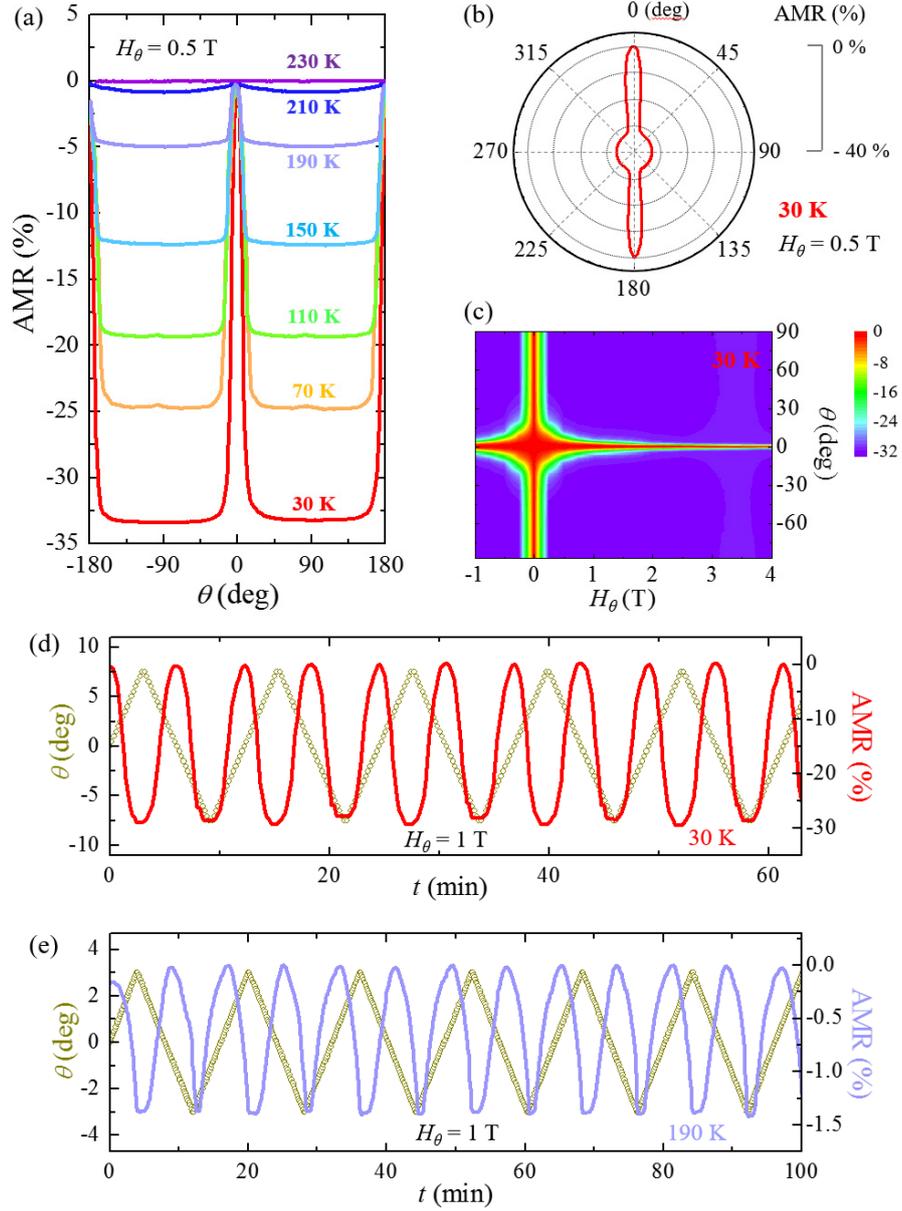

**Figure 3.** (Color online). (a) Angular dependence of the AMR, obtained by rotating $H_\theta = 0.5$ T between $-180°$ and $180°$ at different temperatures from 30 to 230 K. To examine the crystalline component of the AMR, the rotation plane of $H_\theta$ is perpendicular to the current direction. (b) Polar angular plot of the AMR by rotating $H_\theta = 0.5$ T at 30 K. (c) Contour plot of the AMR at 30 K for $H_\theta$ between $-1$ T and 4 T and $\theta$ between $-90°$ and $90°$. (d) Oscillation of AMR at 30 K and $H_\theta = 1$ T, obtained by varying $\theta$ linearly with time between $-7.5$ and $7.5°$. (e) Oscillation of AMR at 190 K and $H_\theta = 1$ T, induced by the linear variation of $\theta$ with time between $-3.0$ and $3.0°$.

# Supporting Information

**Title: Antiferromagnet-based spintronic functionality by controlling isospin domains in a layered perovskite iridate**

*Nara Lee, Eunjung Ko, Hwan Young Choi, Yun Jeong Hong, Muhammad Nauman, Woun Kang, Hyoung Joon Choi\*, Young Jai Choi\*, and Younjung Jo\**

**S1. Macroscopic isospin order in $Sr_2IrO_4$**

$Sr_2IrO_4$ crystallizes in a centrosymmetric tetragonal structure with four-fold rotational symmetry ($C_4$) about the [001] axis[1, 2]. The crystallographic structure is shown in Figure S1 and it was confirmed by a powder X-ray diffractometer (Ultima IV, Rigaku, Tokyo, Japan) using Cu-K$_\alpha$ radiation at room temperature. The detailed characterization of the structure was performed by Rietveld refinement using the FullProf software. The lattice constants were found to be $a$ = 5.490 Å and $c$ = 25.785 Å. Below $T_N$, the spin-orbit entangled $J_{eff}$=1/2 isospins display a canted AFM order in the *ab* plane, where the canting of spins arises from the staggered rotation of $IrO_6$ octahedra. In addition, the rotational symmetry of the system is reduced from $C_4$ to $C_2$, producing an orthorhombic magnetic structure. Therefore, four different magnetic domains are formed under zero magnetic field, as shown in Figure S2 a-d. In more detail, two 180°-oriented magnetic domains in which the ordered magnetic moments are aligned along a unique axis are formed (Figure S2a and b). Orthorhombic twinning is responsible for the other two types of domain (Figure S2c and d). In a magnetic unit cell, the resulting net magnetic moment ($\mu_{net}$) in each $IrO_2$ layer is stacked along the [001] axis, and $\mu_{net}$'s of consecutive $IrO_2$ layers in each unit cell are cancelled out (Figure S2a). To apply a magnetic field above the critical strength in the *ab* plane generates a single weak ferromagnetic state by aligning $\mu_{net}$'s to the field direction (Figure S2e).

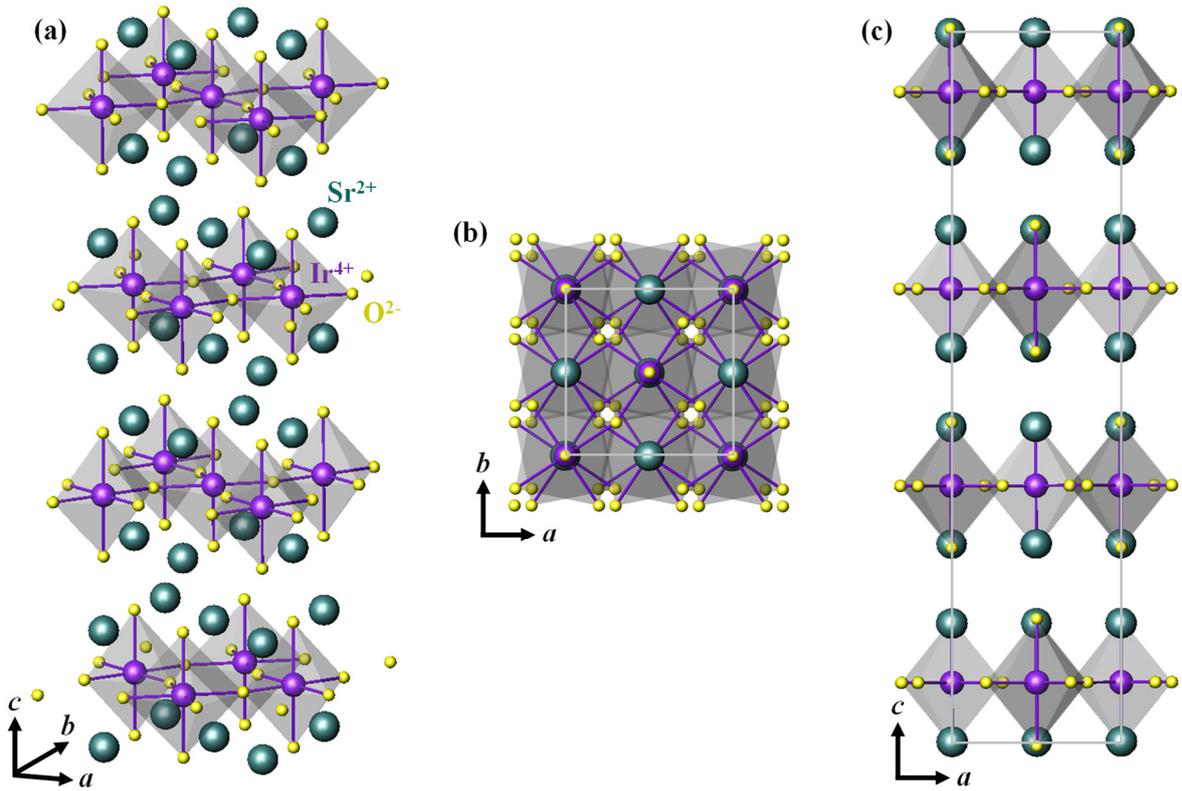

**Figure S1.** (a) Crystallographic structure of $Sr_2IrO_4$. Blue-green, purple, and yellow spheres represent $Sr^{2+}$, $Ir^{4+}$, and $O^{2-}$ ions, respectively. (b) and (c) Views of the crystal structure of layered perovskite $Sr_2IrO_4$ (tetragonal $I4_1/acd$ structure) from the $c$ and $b$ axes, respectively.

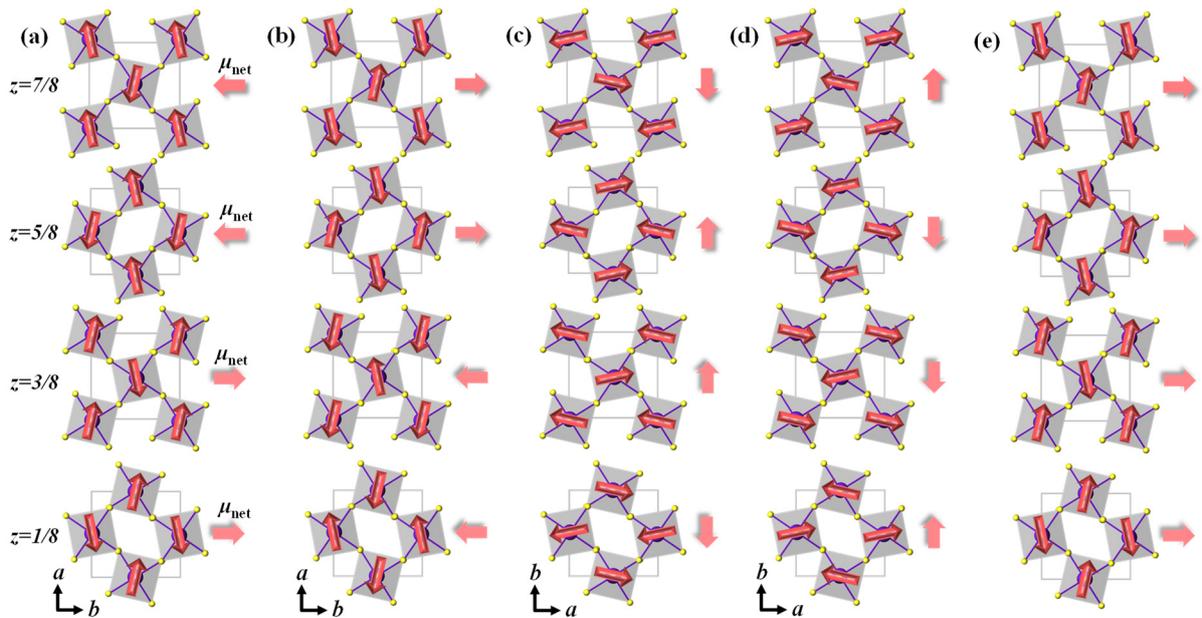

**Figure S2.** (a)-(d) Stacking pattern of the four types of canted AFM domain in $Sr_2IrO_4$. Red arrow to the right of each IrO$_2$ layer denotes the direction of $\mu_{net}$. (e) Stacking pattern of the single weakly ferromagnetic domain in a magnetic unit cell at a magnetic field above the critical strength (~0.3 T) applied in the *ab* plane.

**S2. Band structures associated with canted AFM domains**

To check the influence of the band structures on our large MR effect, we have produced the electronic band structures (Figure S3) associated with the four types of canted AFM domains formed in zero magnetic field and single weakly ferromagnetic state formed in an in-plane magnetic field above the critical strength, illustrated in Figure S2. The band structure in Figure S3a corresponds to either of two 180°-oriented canted AFM domains shown in Figure S2a and b. The band structure in Figure S3b corresponds to either of the two other orthorhombically-twined domains in Figure S2c and d. It appears that the band gaps are all the same for the four types of magnetic domain (Figure S2 a-d). The band structure representing the single weakly ferromagnetic state (Figure S3c) shows that the band gap is 0.0074 eV smaller than that in one of the four types of magnetic domain (Figure S3a or b). The influence of the band structure alteration on our main observation of the angle-sensitive AMR effect can also be specified by comparing band gaps between the four types of canted AFM domain (Figure S2 a-d) sustained when a magnetic field (0.5 T in Figure 3a of the main manuscript) is applied to the hard magnetic axis of [001] direction and the single weakly ferromagnetic state (Figure S2e) formed by deviation of the magnetic field from the [001] axis. Therefore, our estimation manifests that our large MR and angle-sensitive AMR effects cannot be ascribed to the difference in electronic band structures.

Although we used a unit cell containing 8 Ir atoms for Figure S3c, the atomic and magnetic structure of single weakly ferromagnetic state is *I* lattice (body-centered lattice). So we can represent it with a primitive unit cell containing 4 Ir atoms. For more clear presentation and comparison, Figure S3d is drawn using the primitive unit cell containing 4 Ir atoms, showing 4 valence bands and 4 conduction bands in red. The figure of Brillouin zone corresponding to Figure S3d is shown in Figure S3f.

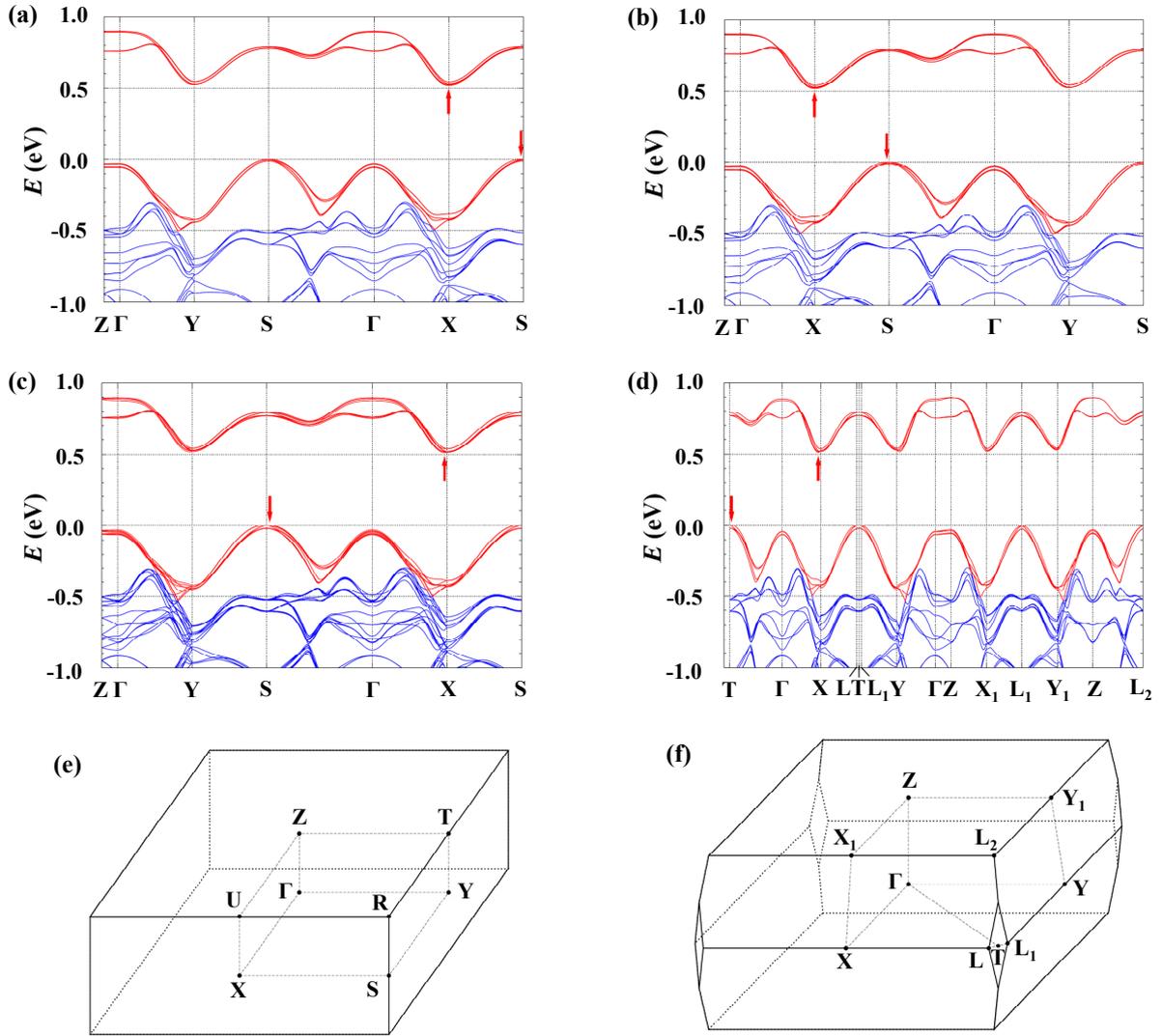

**Figure S3.** (a) Band structure corresponds to either of two 180°-oriented canted AFM domains shown in Figure S2a and b. (b) Band structure corresponds to either of orthorhombically-twined two other domains in Figure S2c and d. The band gaps are all the same for four types of magnetic domain formed in zero magnetic field (Figure S2a-d). (c) Band structure associated with the spin configuration of a single weakly ferromagnetic state in Figure S2e. The red arrows indicate the conduction band minimum (CBM) and valence band maximum (VBM). The CBM is slightly away from X while it is right at X in the other cases ((a) and (b)). (d) Band structure associated with the spin configuration of a single weakly

ferromagnetic state in the primitive unit cell containing 4 Ir atoms. (e) Brillouin zone associated with band structures of a-c. (f) Brillouin zone with band structure of d.

## S3. Formation of $J_{eff}$ = 1/2 isospin state in $Sr_2IrO_4$

In $Sr_2IrO_4$, $Ir^{4+}$ 5d orbitals are split to $t_{2g}$ and $e_g$ orbitals due to crystal splitting, and electronic states near the Fermi energy are mostly of $t_{2g}$ orbitals. For the three-fold $t_{2g}$ orbitals, one can introduce an effective orbital angular momentum of $L_{eff}$ = 1 by regarding $|m_{l_{eff}} = 0\rangle = |d_{xy}\rangle$ and $|m_{l_{eff}} = \pm 1\rangle = \mp\frac{1}{\sqrt{2}}(|d_{yz}\rangle \pm i|d_{xz}\rangle)$ as eigenstates of $L_{eff,z}$ corresponding to eigenvalues of 0 and $\pm\hbar$, respectively[3, 4]. Then one obtains $\vec{L}_{eff} = -\vec{L}$ so that the spin-orbit coupling (SOC) can be expressed with the Hamiltonian $H_{soc} = \alpha\vec{L}\cdot\vec{S} = -\alpha\vec{L}_{eff}\cdot\vec{S} = -\frac{\alpha}{2}(J_{eff}^2 - L_{eff}^2 - S^2)$, where $\vec{S}$ is the electron spin and $\vec{J}_{eff}$ is the effective total angular momentum defined as $\vec{J}_{eff} = \vec{L}_{eff} + \vec{S}$. Thus, in the strong SOC limit, the $t_{2g}$ bands are split into four $J_{eff}$=3/2 bands and two $J_{eff}$=1/2 bands. In $Sr_2IrO_4$, the $J_{eff}$=3/2 bands are mostly below the Fermi energy and the $J_{eff}$=1/2 bands are right at the Fermi energy.

As shown in Figure S4a which was obtained using the local density approximation (LDA) with SOC, the two $J_{eff}$=1/2 bands are metallic without an energy gap unless $Ir^{4+}$ ions have magnetic moment. The Coulomb interaction and AFM order together with SOC result in splitting of the $J_{eff}$ = 1/2 bands with an energy gap (see Figure 2a in the main manuscript). We note that the $Ir^{4+}$ moment at each atomic site, $\vec{M}_{Ir} = \sum_{occ}\frac{-|e|}{2m}\langle\vec{L} + 2\vec{S}\rangle$, is parallel to the $J_{eff}$=1/2 isospin of occupied states, i.e., $\sum_{occ}\langle\vec{J}_{eff}\rangle$ because $\langle\vec{S}\rangle = -\frac{\hbar}{6}\vec{\sigma}$, $\langle\vec{L}\rangle = -\frac{2\hbar}{3}\vec{\sigma} = -\langle\vec{L}_{eff}\rangle$, and $\langle\vec{J}_{eff}\rangle = \langle\vec{L}_{eff} + \vec{S}\rangle = \frac{\hbar}{2}\vec{\sigma}$, where $\vec{\sigma}$ denotes the Pauli matrices. Since the arrangement of isospins in each $IrO_2$ plane is represented by the direction of $\mu_{net}$, the rotation of $\mu_{net}$'s determines the characteristic of a canted AFM domain. As shown in Figure 2b and c (main manuscript), this locking of the isospin to the direction of the $\mu_{net}$ is strong for the bands near the X-S line while the isospin moment ($\langle \vec{J}_{\text{eff}} \rangle$) is weak near the Γ point due to hybridization between $t_{2g}$ and $e_g$ bands. As the conduction band minimum (CBM) is at the X point and the valence band maximum (VBM) is at the S point as shown in Figure 2a (main manuscript), each charge carrier has the well-defined isospin whose direction is controlled by the direction of the $\mu_{\text{net}}$.

In Sr$_2$IrO$_4$, the characteristic of a wave function in CBM or VBM is well represented by the arrangement of isospins, the direction of the $\mu_{\text{net}}$ has a strong influence on the character of the wave function. As the $\mu_{\text{net}}$ is rotated by an angle $\theta$, all the isospins within the IrO$_2$ plane are also rotated by the same angle. Since the isospin direction is locked to the $\mu_{\text{net}}$, our interest then is to calculate the squared inner product of a state and its transformed state after the rotation of the $\mu_{\text{net}}$. Similarly to the case of the real electron spin, if two states $\psi_1$ and $\psi_2$ have isospins with an angle $\theta$, their squared inner product $|\langle\psi_1|\psi_2\rangle|^2$ is given by $|\langle\psi_1|\psi_2\rangle|^2 = \sin^2\theta$. Then, $|\langle\psi_1|\psi_2\rangle|^2 = \frac{1}{2}$ for $\theta$ = 90° and $|\langle\psi_1|\psi_2\rangle|^2 = 0$ for $\theta$ = 180°. In the case of rotating the $\mu_{\text{net}}$ by 180°, the collective rotation of the isospins is associated with time-reversal-symmetry transformation of the wave function which is completely orthogonal to that before the transformation. As a result, the squared inner products of wave functions before and after the rotation leads to the complete reflection of the electrical current at the domain boundary. Because our unit cell of Sr$_2$IrO$_4$ contains 8 Ir$^{4+}$ ions, our band structure from the LSDA+U method with SOC has eight almost-degenerate $J_{\text{eff}}$ = 1/2 conduction bands and eight almost-degenerate $J_{\text{eff}}$ = 1/2 valence bands. To avoid mapping of individual states before and after the rotation of the $\mu_{\text{net}}$, we instead evaluated $f_k^{CB}(\theta) = \frac{1}{8}\sum_{n=1}^{8}\sum_{m=1}^{8}\left|\langle\psi_{n,k}^{CB}|\psi_{m,k}^{CB,\theta}\rangle\right|^2$ and $f_k^{VB}(\theta) = \frac{1}{8}\sum_{n=1}^{8}\sum_{m=1}^{8}\left|\langle\psi_{n,k}^{VB}|\psi_{m,k}^{VB,\theta}\rangle\right|^2$, where $\psi_{n,k}^{CB}$ and $\psi_{n,k}^{VB}$ are conduction band (CB) and valence band (VB) states, respectively, at the crystal momentum $k$ before the rotation of the $\mu_{\text{net}}$, and $\psi_{n,k}^{CB,\theta}$ and $\psi_{n,k}^{VB,\theta}$ are corresponding states after the rotation of the $\mu_{\text{net}}$ by an angle $\theta$. Figure 2d and e (main manuscript) and Figure S4b and c show $f_k^{CB}(\theta = 90°)$, $f_k^{CB}(\theta = 180°)$, $f_k^{VB}(\theta = 90°)$, and $f_k^{VB}(\theta = 180°)$, respectively. These figures show that $f_k^{CB}(\theta = 90°) \approx 0.5$, $f_k^{CB}(\theta = 180°) \approx 0$, $f_k^{VB}(\theta = 90°) \approx 0.5$, and $f_k^{VB}(\theta = 180°) \approx 0$ near the M-X line, confirming that the isospin $J_{eff} = 1/2$ is an effectively good quantum number and the electronic transport at domain boundaries is determined by the characteristics of wave functions before and after rotating the $\mu_{net}$.

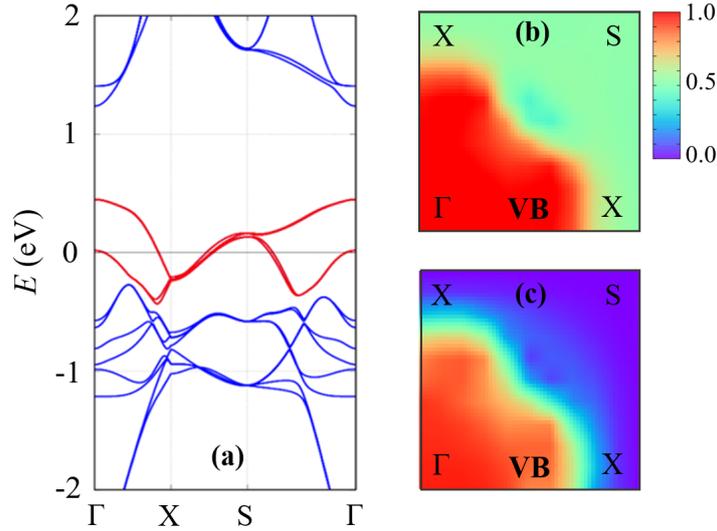

**Figure S4.** (a) Band structure calculated from the LDA+SOC method neglecting $Ir^{4+}$ magnetic moment. (b)-(c) Squared inner products of VB states before and after the rotation of $\mu_{net}$ by 90° (b) and 180° (c).

## S4. Transport properties modelled by microscopic theory

To examine the strong temperature dependence of the AMR, we devise a microscopic theory for the electron transport. First, as described above, electrons in our sample experience scattering due to isospin mismatch as they move across the magnetic domain boundaries. Because the typical size of the magnetic domain depends very weakly on temperature below $T_N$, the scattering rate due to the isospin mismatch ($\Gamma_{DB}(H)$) also depends very weakly on temperature. In contrast, in-plane magnetic field affects $\Gamma_{DB}(H)$ strongly as it eliminates the domain boundaries. Thus, we regard that $\Gamma_{DB}(H)$ depends only on the magnetic field, becoming zero above the critical magnetic field. Second, considering the observed saturation of the $MR_{90}$ in sufficient magnetic field that saturated the $M_{90}$, we conclude that electron transport is affected negligibly by the field except for scattering at domain boundaries. Thus, we regard that the scattering rate inside each magnetic domain ($\Gamma_0(T)$) does not depend on the magnetic field, and thus is only dependent on temperature. Finally, we consider that our sample has defects, such as oxygen deficiencies, that can donate electrons to the CB. With these considerations, we can express the resistivity of our sample as $\rho(T,H) = \{\frac{m_c}{n_c(T)e^2} + \frac{m_v}{n_v(T)e^2}\}\{\Gamma_0(T) + \Gamma_{DB}(H)\}$. Here, $n_c(T)$ and $n_v(T)$ are the electron and hole densities at the CB and VB, respectively, while $m_c$ and $m_v$ are the effective electron and hole masses at the CBM and VBM, correspondingly. Because $\Gamma_{DB}(H)$ vanishes above the critical magnetic field of $H_c \approx 0.3$ T, the field-induced decrease in the resistivity is expressed by $\Delta\rho(T) = \rho(T, H > H_c) - \rho(T, H = 0) = \{\frac{m_c}{n_c(T)e^2} + \frac{m_v}{n_v(T)e^2}\}\Gamma_{DB}(H = 0)$, which is determined only by the thermal distribution of charge carriers in the band structure and $\Gamma_{DB}(H)$ in zero magnetic field. We introduce a Gaussian density of states (DOS) for the defect level inside the energy gap (Figure S5a) and fit the temperature-dependent $|\Delta\rho(T)|$ as shown in Figure S5b, obtaining the defect concentration of $1.4\times10^{19}$ m$^{-3}$, the defect-level binding energy of 0.22 eV below the CBM, the defect-level broadening (i.e., standard deviation) of 0.052 eV, and the scattering rate $\Gamma_{DB}(H = 0)$ of $1.8\times10^7$ s$^{-1}$. Figure S6b also shows the maximum AMR value at each temperature, i.e., AMR$_{Max}$ (%) = $\frac{R(90°)-R(0°)}{R(0°)} \times 100$. Below $T_N \approx 220$ K, AMR$_{Max}$ depends linearly on the temperature and exhibits the largest magnitude of ~33% at 30 K. The emergence of the AMR effect coincides with the onset of the canted AFM order, supporting the strong interconnection between the AMR and isospin order. As the defect level is closer in energy to the CBM than to the VBM, $n_c(T)$ is greater than $n_v(T)$ at low temperatures (Figure S5c). From these results and the measured $\rho(T, H = 0)$, we estimate $\Gamma_0(T)$ in the interior of each domain, which is observed to increase with temperature (Figure S6d). A polynomial fit shows that $\Gamma_0(T) = \Gamma_0(T = 0) + aT^5$, where $\Gamma_0(T = 0) = 5.2\times10^7$ s$^{-1}$ and $a = 0.0011$ s$^{-1}$K$^{-5}$. The temperature-independent term is typically from defect scattering, while the $T^5$ term suggests scattering by acoustic phonons with small momenta as in metals[5].

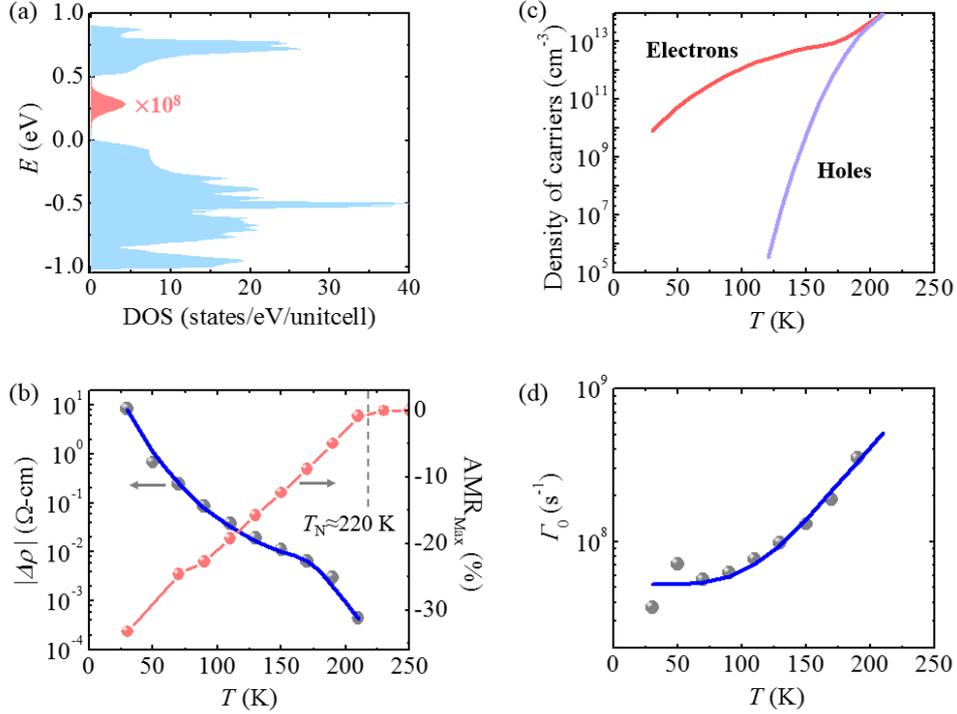

**Figure S5.** (a) Calculated DOS of a $Sr_2IrO_4$ crystal with defects. The DOS of a perfect $Sr_2IrO_4$ crystal appears light blue while the defect-induced DOS used for transport calculation is illustrated in pink, magnified by $10^8$. (b) Difference between the resistivities measured at $H_{90}$ = 0 T and $H_{90}$ = 0.5 T. Grey dots represent the experimental result and the blue line represents the theoretical result based on the DOS shown in (a) and the temperature-independent $\Gamma_{DB}(H=0) = 1.8\times10^7$ s$^{-1}$ at the magnetic domain boundaries. The maximum AMR, extracted from the data of Figure 3a (main manuscript), is plotted in lined red dots. (c) The $n_c(T)$ in the CB and the $n_v(T)$ in the VB calculated from the DOS shown in (a). (d) $\Gamma_0(T)$ in the interior of the magnetic domains, estimated from our experimental results and theoretical model calculation and shown in logarithmic scale. The results (grey dots) are well fitted by the polynomial curve (blue line) $\Gamma_0(T) = \Gamma_0(T=0) + aT^5$, where $\Gamma_0(T=0) = 5.2\times10^7$ s$^{-1}$ and $a$ = 0.0011 s$^{-1}$K$^{-5}$.

**S5. Measurement of crystalline AMR component by rotating field in the *ab* plane perpendicular to current along the [001] axis**

The AMR in phenomenology involves two different components: a noncrystalline component resulting from the orientation of magnetization relative to a specific current direction, and a crystalline component arising entirely from crystal symmetry[6, 7]. In pursuance of the AMR effect driven by magnetocrystalline anisotropy, the measurements of the crystalline AMR component for single-crystalline bulk and thin film $Sr_2IrO_4$ samples were previously reported[8, 9]. The AMR was measured with rotation of $H_\varphi$ in the *ab* plane perpendicular to current along the [001] axis, where $\varphi$ is the angle deviating from the [100] axis in the *ab* plane (inset of Figure S6a). In a 6-nm-thick $Sr_2IrO_4$ thin film, the rotation of $IrO_6$ octahedra disappeared with lowering the Nèel temperature down to ~100 K[9]. Thus, $Sr_2IrO_4/La_{2/3}Sr_{1/3}MnO_3/SrTiO_3$ thin film structure was prepared and the AMR effect with four-fold rotational symmetry was observed within 1% variation of resistivity by rotating the AFM moment in $Sr_2IrO_4$, dragged by the ferromagnetic moment in $La_{2/3}Sr_{1/3}MnO_3$ via the exchange spring effect[9]. The small and sinusoidal variation of the AMR effect in thin films was ascribed to anisotropic electronic structure depending on the spin axis. The calculation of electronic structure depending on the spin axis resulted in a band gap difference between the [100] and [110] axes of 0.03 eV[9]. However, as explained in S2, the estimated band gap difference of 0.0074 eV relevant to our main observation of the AMR effect strongly suggests that the highly anisotropic nature of our MR data cannot be explained by the anisotropic electronic structure model. The AMR measurement using point-contact geometry for the bulk specimen revealed the crossover from four-fold to two-fold rotational symmetry upon increasing magnetic fields[8]. The mechanism for the effect was speculated as magneto-elastic alteration of bond angles in the $IrO_6$ octahedra[8]. In Figure S6a, the resistivity in our bulk $Sr_2IrO_4$, measured in zero magnetic field with current along the [001] axis, exhibits a semiconducting temperature dependence without any anomaly at $T_N$ of the resistivity. For comparison with the previous results, we also measured the bulk

AMR, defined as AMR (%) = $\frac{R(\varphi) - R(0°)}{R(0°)} \times 100$ at 120 K (Figure S6b). The observed AMR reveals the highly harmonic angular dependence of $\cos 4\varphi$ with about 3% change of resistivity in $H_\varphi$ = 1 T. The four-fold rotational symmetry of the AMR is manifestly shown in the polar plot of Figure S6c. Different from the point-contact measurement in the previous work[18], we only observe the four-fold rotational symmetry because all the spin easy axes are equivalent at the macroscopic scale because of the orthorhombic twinning.

To examine the contribution of noncrystalline component to the AMR in our bulk $Sr_2IrO_4$, the AMR was measured by rotating $H_\varphi$ = 1 T in the *ab* plane with current along the [110] axis at 60 K. A conventional AMR effect in polycrystalline 3*d* ferromagnetic metals is attributed predominantly to the noncrystalline AMR component with $\cos 2\varphi$ dependence [6]. However, Figure S7 shows that the AMR is proportional to the highly harmonic component of $\cos 4\varphi$ in which the four-fold rotational symmetry is maintained although the orientation of $H_\varphi$ was relative to the current direction during the measurement. The result suggests that the contribution of noncrystalline component to the AMR is much weaker than that of crystalline component in $Sr_2IrO_4$.

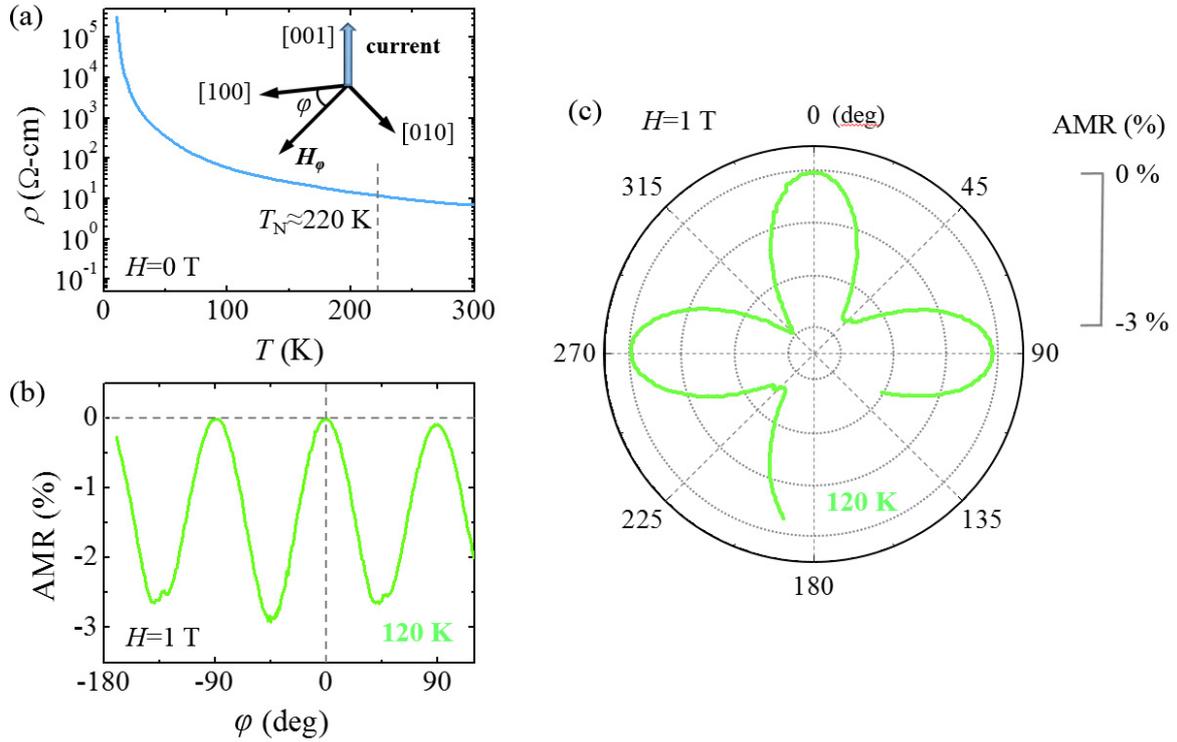

**Figure S6.** (a) Temperature dependence of resistivity measured with current along the [001] axis and zero magnetic field. Inset shows the geometry of resistivity and AMR measurements; $\varphi$ is the angle of rotation with respect to the [100] axis in the $ab$ plane, and the rotation plane is perpendicular to the current direction. (b) Angular dependence of the AMR at $H = 1$ T and $T = 120$ K. (c) Polar plot of the AMR at $H = 1$ T and $T = 120$ K.

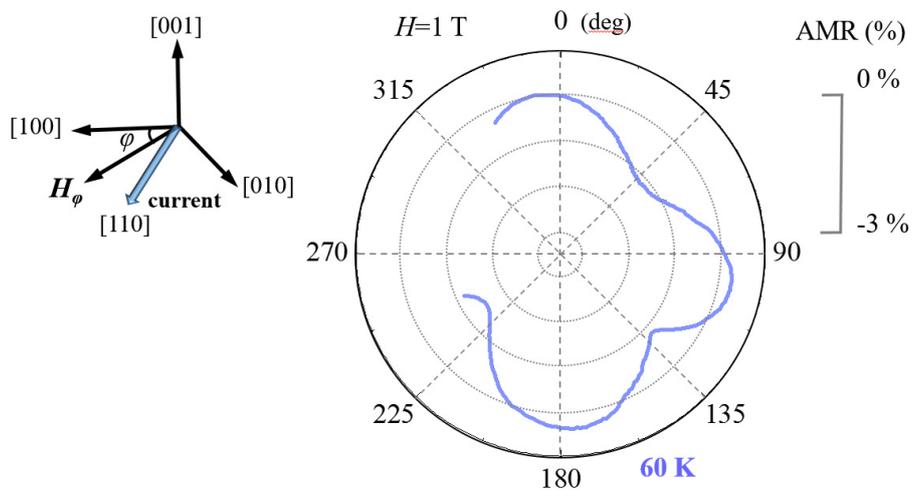

**Figure S7.** Polar plot of noncrystalline component of AMR at $H = 1$ T and $T = 60$ K. Inset shows the geometry of AMR measurement; $\varphi$ is the angle of rotation with respect to the [100] axis in the $ab$ plane.